\documentclass[prl,aps,
amssymb,
twocolumn,
floatfix,
preprintnumbers,
superscriptaddress,longbibliography,10pt]{revtex4-1}

\usepackage[dvipsnames]{xcolor}
\usepackage{amsmath,amssymb,mathrsfs,amsthm,tikz,shuffle,paralist,accents}
\usepackage{subcaption}
\usepackage{enumitem}
\usepackage{mathtools}
\usepackage{slashed}
\usepackage{hyperref}
\usepackage{cleveref}

\newcommand{\myI}[1]{{\rm I}_{[#1]}}
\newcommand{\mydeg}[1]{\deg\{#1\}}
\allowdisplaybreaks
\begin{document}

\title{Taming Symbolic IBP Reduction with Intermediate Bases}

\author{Qian Song}
 \email{qian.song@ugent.be}
\affiliation{%
Department of Physics and Astronomy, Ghent University, 9000 Ghent, Belgium
}%

\begin{abstract}

Despite many years of development in integration-by-parts reduction, reconstructing all reduction coefficients, which are rational functions of kinematic variables and the space-time dimension, remains a non-trivial problem. The main difficulty comes from the large number of unknowns in a general ansatz, which can lead to a linear system that is too large to solve. In this paper, we present an algorithm for reconstructing reduction coefficients through a sequence of intermediate bases. The resulting analytic reduction coefficients are products of a few analytic matrices, whose non-zero entries are simple rational functions. We demonstrate the efficiency of this algorithm with two cutting edge examples: a three-point massive box-triangle and a four-point massive pentagon-triangle. Reconstructing all reduction coefficients for the box-triangle (pentagon-triangle) requires 3289 (13013) numerical samplings, significantly fewer than the number of unknowns in the general ansatz, 1407406 (21638331).

\end{abstract}

\maketitle


\section{Introduction}
In the modern computational framework for scattering amplitudes, an essential step is to reduce all Feynman integrals to a basis of master integrals \cite{Chetyrkin:1981qh,Laporta:2000dsw} using integration-by-parts (IBP) reduction. Many public packages exist \cite{Anastasiou:2004vj,Lee:2012cn,Lee:2013mka,Smirnov:2008iw,Smirnov:2013dia,Smirnov:2014hma,Smirnov:2019qkx,Studerus:2009ye,vonManteuffel:2012np,Maierhofer:2017gsa,Klappert:2020nbg,Wu:2023upw}, but multi-scale IBP reduction remains a major bottleneck. Direct symbolic reduction suffers from severe expression swell and, for frontier problems, the simplification of intermediate coefficients can dominate the run time\cite{Smirnov:2024sfi}. A common alternative is to perform numerical IBP reduction, often over finite fields, and reconstruct the analytic reduction coefficients from sufficiently many samplings \cite{vonManteuffel:2014ixa,Peraro:2016wsq,Klappert:2019emp,Peraro:2019svx,Klappert:2020nbg,Maier:2024djk,Belitsky:2023qho,10.1145/1005285.1005321,10.1145/62212.62241,zippel_1989,gdmm1997,10.1145/42267.45069,ket1988,10.1145/1277548.1277577,10.1145/1073884.1073903,10.1145/281508.281538,10.1145/345542.345629,huang2017sparsepolynomialinterpolationfinitely,Klappert:2020aqs,Floater:2007bri}. 

Reconstructing reduction coefficients are useful not only for obtaining analytic scattering amplitudes, but also for building differential equations for master integrals, identifying spurious singularities, and enabling fast numerical evaluation \cite{Henn:2013pwa,Lee:2014ioa}. However, without any prior knowledge of the analytic structure of the coefficients, the number of required samplings can be enormous, making the reconstruction itself a difficult computational problem.

Several ideas have been developed to reduce the computational cost of reconstruction: one can exploit linear or polynomial
relations among reduction coefficients \cite{Liu:2023cgs}, make a judicious choice of master integrals that factorize the dependence in kinematic variables and the space-time dimension $D$ \cite{Smirnov:2020quc,Usovitsch:2020jrk}, and reduce the number of unknowns in the ansatz by identifying denominator factors or partial fraction forms \cite{Maier:2024djk, Abreu:2018zmy,DeLaurentis:2022otd,Chawdhry:2023yyx,Smirnov:2026kjb}. 

In this paper, we propose a new approach of reconstructing reduction coefficients. This approach is motivated by the observation that the IBP relations from which the reduction starts usually have simple polynomial coefficients, even when the final reduction coefficients are complicated. The basic idea of this approach to express complicated Feynman integrals in terms of simpler ones, which we call an intermediate basis, by row reducing a smaller set of IBP relations. These reduction rules are then inserted back into the original IBP system, so that the complicated integrals no longer appear. Iterating this procedure eventually expresses all integrals in terms of the master integrals. Instead of one transformation matrix with complicated entries, all reduction coefficients are represented as a product of a few sparse matrices whose entries are simple rational functions, namely
\begin{align}
&\begin{pmatrix}
J_1 \\ \vdots \\ J_m
\end{pmatrix} = \overline{M}_{m\times n}\begin{pmatrix}
J_{m-n+1} \\ \vdots \\ J_m
\end{pmatrix} \nonumber \\
&=\underbrace{M_{m\times (m-n_1)} \times \cdots \times M_{(n+n_k)\times n}}_{k~{\rm times}} \begin{pmatrix}
J_{m-n+1} \\ \vdots \\ J_m
\end{pmatrix} \label{eq:mat}
\end{align}
with $n=m-n_1-\cdots-n_k$. The set $\{J_1,\cdots,J_m\}$ is the ordered list of Feynman integrals appearing in the IBP relations, and the last $n$ elements, $\{J_{m-n+1},\cdots,J_m\}$ are the master integrals. The reduction result shown in \cref{eq:mat} uses $k-1$ intermediate bases, $\{J_{n_1+\cdots+n_i+1}, \cdots, J_m\}$ with $i=1,\dots,k-1$.  One can clearly see that $i-$th intermediate basis removes the first $n_i$ integrals and the reduction coefficients are expressed as products of $k$ basis transformation matrices.

The representation in \cref{eq:mat} has two advantages. First, reconstructing the individual basis transformation matrices requires significantly fewer samples than directly reconstructing $\overline{M}$. The required number of samples is bounded by the most complicated entry among the basis transformation matrices, rather than by the most complicated entry of their product. This bound can be tuned by the choice of intermediate bases. Second, the product form gives a compact analytic representation that is well suited for fast numerical evaluation, since complicated rational functions are decomposed into simpler pieces. 

The paper is organized as follows. In Sec.\ref{sec:method} we introduce the algorithm for finding intermediate bases, starting with an intuitive example. In Sec.\ref{sec:example}, we apply our approach to two Feynman integral families 
relevant to two-loop electroweak corrections to $e^+e^-\to {\rm ZH}$
. Conclusion is given in Sec.\ref{sec:conclusion}.

\section{Method}\label{sec:method}
\subsection{Intuitive Example} \label{sec:intExample}
\begin{figure}
    \centering
    \includegraphics[width=0.7\linewidth]{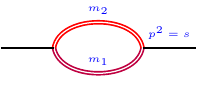}
    \caption{1-loop massive bubble with non-equal mass.}
    \label{fig:bubble}
\end{figure}

For an intuitive understanding of the approach, we use the one-loop massive bubble as an example.  The corresponding Feynman diagram is shown in Fig.\ref{fig:bubble}. We want to reduce the following integrals 
\begin{align}
\{ \myI{2,2}, \myI{2,1}, \myI{1,2}, \myI{2,0}, \myI{0,2} \} \label{eq:bubbleFIs}
\end{align}
into the master integrals $\{ \myI{1,1}, \myI{1,0}, \myI{0,1}\}$. The lower indices denote the exponents of the propagators. The five IBP relations are generated by NeatIBP\cite{Wu:2023upw}, which reads   
\begin{align}
0&= \lambda \myI{2,2} + ((D-4)\delta_-+\delta_+)\myI{2,1}+\delta_+\myI{1,2}\nonumber \\
&+(4-D)\myI{2,0} \label{eq:bubbleIBP1} \\ 
0&=2m_1^2\myI{2,1}+\delta_+\myI{1,2}+(3-D)\myI{1,1} +\myI{0,2}  \label{eq:bubbleIBP2} \\ 
0&=\lambda \myI{1,2} + (D-3)\delta_-\myI{1,1} +\delta_+\myI{0,2} \nonumber\\
&+(2-D) \myI{1,0} \label{eq:bubbleIBP3} \\  
0&= 2m_1^2 \myI{2,0} +(2-D) \myI{1,0} \label{eq:bubbleIBP4}\\ 
0&= 2m_2^2 \myI{0,2} +(2-D) \myI{0,1}  \label{eq:bubbleIBP5}
\end{align} 
where $\lambda=(m_1^2+m_2^2+s)^2-2m_1^2s-2m_2^2s-2m_1^2m_2^2,\delta_+=m_1^2+m_2^2-s,\delta_-=m_1^2-m_2^2+s$. Since the size of the IBP system is small, we can directly perform symbolic row reduction. The reduction coefficients of \cref{eq:bubbleFIs} are
\begin{align}
\myI{2,2}&=c_{0,1}\dfrac{D-2}{2m_2^2\lambda^2} \myI{0,1}+ c_{0,1}|_{m_1\leftrightarrow m_2}  \dfrac{D-2}{2m_1^2\lambda^2} \myI{1,0} \nonumber \\
&+(c_{0,1}+(D-5)\delta_+\delta_-)\dfrac{3-D}{\lambda^2}\myI{1,1} \label{eq:bubbleres1}\\
\myI{1,2}&=\dfrac{D-2}{\lambda}\myI{1,0} + \dfrac{(2-D)\delta_+}{2m_2^2\lambda}\myI{0,1} \nonumber\\
&+ \dfrac{(3-D)\delta_-}{\lambda} \myI{1,1} \label{eq:bubbleres2}\\
\myI{2,1}&=\myI{1,2}|_{m_1\leftrightarrow m_2, \myI{1,0}\leftrightarrow \myI{0,1}} \label{eq:bubbleres3}\\
\myI{0,2}&=\dfrac{D-2} {2m_2^2}\myI{0,1},\label{eq:bubbleres4}\\
\myI{2,0}&=\myI{0,2}|_{m_1\leftrightarrow m_2,\myI{1,0}\leftrightarrow \myI{0,1}} \label{eq:bubbleres5}
\end{align}
with $c_{0,1}=2(5-D)m_2^2\delta_-+\lambda$. 

Throughout this paper, all reduction coefficients are regarded as rational functions of the independent kinematic variables $\vec{s}$ and the space-time dimension $D$; for the one-loop bubble, $\vec{s}=\{m_1^2,m_2^2,s\}$. We write a rational function as $r=p/q$ with $p,q$ coprime and use a pair of integers to denote the degree of $r$,
\begin{align}
\deg r = \mydeg{d_1,d_2}
\end{align}
where $d_{1}$ and $d_2$ represent the degree of $p$ and $q$, respectively. Explicitly, 
\begin{align}
p = \sum_i \kappa_i D^{\beta_i} \prod_j  s_j^{\alpha_{i,j}},~d_1 = \max_i\Big(\sum_j \alpha_{i,j}+\beta_i\Big),
\end{align} 
with $\kappa_i\in \mathbb{Q}, \alpha_{i,j}\in \mathbb{Z}_{\geq 0}, \beta_i\in \mathbb{Z}_{\geq 0}$. The degree of the denominator polynomial $q$ is defined similarly. Note that $p,q$ have fixed mass dimensions but are not necessarily homogeneous in $D$ and $\vec{s}$, so $d_1,d_2$ are the maximum degrees among all monomials.

To compare the complexity of two rational functions, let
\begin{align}
\deg(r_1)=\mydeg{a_1,a_2},~
\deg(r_2)=\mydeg{b_1,b_2}.
\end{align}
We say that $r_1$ is more complicated than $r_2$ if
\begin{align}
\{a_1+a_2,a_1,a_2\}>\{b_1+b_2,b_1,b_2\}
\end{align}
where $>$ is understood lexicographically.

Following such notation, one can clearly see that the most complicated reduction coefficients has degree $\mydeg{4,5}$, which comes from the reduction coefficients in $\myI{2,2}$. While the polynomial coefficients in \cref{eq:bubbleIBP1,eq:bubbleIBP2,eq:bubbleIBP3,eq:bubbleIBP4,eq:bubbleIBP5} are rather simple, which are at most $ \mydeg{2,0}$. It is therefore natural to use \cref{eq:bubbleIBP1} to rewrite $\myI{2,2}$ in terms of other Feynman integrals. Explicitly,
\begin{align}
\myI{2,2}&=\dfrac{(4-D)\delta_--\delta_+}{\lambda}\myI{2,1}-\dfrac{\delta_+}{\lambda}\myI{1,2} \nonumber\\
&+\dfrac{D-4}{\lambda}\myI{2,0}  \label{eq:bubbleSol1}
\end{align}
Plugging \cref{eq:bubbleSol1} into \cref{eq:bubbleIBP1,eq:bubbleIBP2,eq:bubbleIBP3,eq:bubbleIBP4,eq:bubbleIBP5}, we obtain a new set of IBP relations in terms of all Feynman integrals except $\myI{2,2}$. Explicitly, \cref{eq:bubbleIBP1} vanishes, and the rest don't change. Similarly, we use \cref{eq:bubbleIBP2} to rewrite $\myI{2,1}$
\begin{align}
\myI{2,1}&=-\dfrac{\delta_+}{2m_1^2}\myI{1,2} + \dfrac{D-3}{2m_1^2}\myI{1,1}-\dfrac{1}{2m_1^2}\myI{0,2} \label{eq:bubbleSol2}
\end{align}
Again, applying this replacement rules makes \cref{eq:bubbleIBP2} vanish. We left with three IBP relations \cref{eq:bubbleIBP3,eq:bubbleIBP4,eq:bubbleIBP5}. Applying row reduction leads to \cref{eq:bubbleres2,eq:bubbleres4,eq:bubbleres5}.

Recasting reduction rules \cref{eq:bubbleSol1,eq:bubbleSol2,eq:bubbleres2,eq:bubbleres4,eq:bubbleres5} into \cref{eq:mat} gives two intermediate bases
\begin{align}
 \{ \myI{2,1},\myI{1,2},\myI{2,0},\myI{0,2},\myI{1,1},\myI{1,0},\myI{0,1}\} \\
\{ \myI{1,2},\myI{2,0},\myI{0,2},\myI{1,1},\myI{1,0},\myI{0,1}\}
\end{align}
and three transformation matrices 
\begin{align}
M_{8\times 7} &=\begin{pmatrix}
\dfrac{(4-D)\delta_--\delta_+}{\lambda} & -\dfrac{\delta_+}{\lambda}  & \dfrac{D-4}{\lambda} & 0 & 0 & 0 & 0 \\
1 & 0 & 0 & 0 & 0 & 0 & 0\\
0 & 1 & 0 & 0 & 0 & 0 & 0\\
0 & 0 & 1 & 0 & 0 & 0 & 0\\
0 & 0 & 0 & 1 & 0 & 0 & 0\\
0 & 0 & 0 & 0 & 1 & 0 & 0\\
0 & 0 & 0 & 0 & 0 & 1 & 0\\
0 & 0 & 0 & 0 & 0 & 0 & 1\\
\end{pmatrix} \\
M_{7\times 6} &= \begin{pmatrix}
-\dfrac{\delta_+}{2m_1^2} & 0 & -\dfrac{1}{2m_1^2} &\dfrac{D-3}{2m_1^2} & 0 & 0\\
1 & 0 & 0 & 0 & 0 & 0\\
0 & 1 & 0 & 0 & 0 & 0\\
0 & 0 & 1 & 0 & 0 & 0\\
0 & 0 & 0 & 1 & 0 & 0\\
0 & 0 & 0 & 0 & 1 & 0\\
0 & 0 & 0 & 0 & 0 & 1\\
\end{pmatrix} \\
M_{6\times 3} &=\begin{pmatrix}
\dfrac{(3-D)\delta_-}{\lambda} & \dfrac{D-2}{\lambda} & \dfrac{(2-D)\delta_+}{2m_2^2\lambda} \\
0 & \dfrac{D-2}{2m_1^2} & 0 \\
0 & 0 & \dfrac{D-2}{2m_2^2} \\
1 & 0 & 0\\
0 & 1 & 0\\
0 & 0 & 1\\
\end{pmatrix}
\end{align}
One can verify that $M_{8\times 7}\times M_{7\times 6} \times M_{6\times 3}$ leads to the same reduction coefficients as \cref{eq:bubbleres1,eq:bubbleres2,eq:bubbleres3,eq:bubbleres4}, while reconstructing reduction coefficients requires fewer samplings since the most complicated matrix element is $\mydeg{2,3}$. 

The reduction process shown above can be summarized as follows: we iteratively derive reduction rules by row reduction, expressing a subset of complicated Feynman integrals in terms of simpler ones, until the final reduction rules only depend on the master integrals. Although the row reduction was perform symbolically in last section, in practice we have to set kinematic variables and $D$ to random numerical values since symbolic row reduction of large matrix is computationally heavy. 

The complexity of a Feynman integral is determined by the degree of the most complicated rational function in its reduction coefficients. For example, \cref{eq:bubbleres1,eq:bubbleres2,eq:bubbleres3,eq:bubbleres4,eq:bubbleres5} imply the ordering
\begin{align}
\myI{2,2} \succ \myI{2,1} \sim \myI{1,2} \succ \myI{0,2} \sim \myI{2,0} 
\end{align}
and we put the master integrals in the end. These degrees can be simply obtained by reconstructing all reduction coefficients on a univariate slice of kinematic variables and $D$. 

\subsection{The Algorithm}
\begin{figure}
    \centering
    \includegraphics[width=\linewidth]{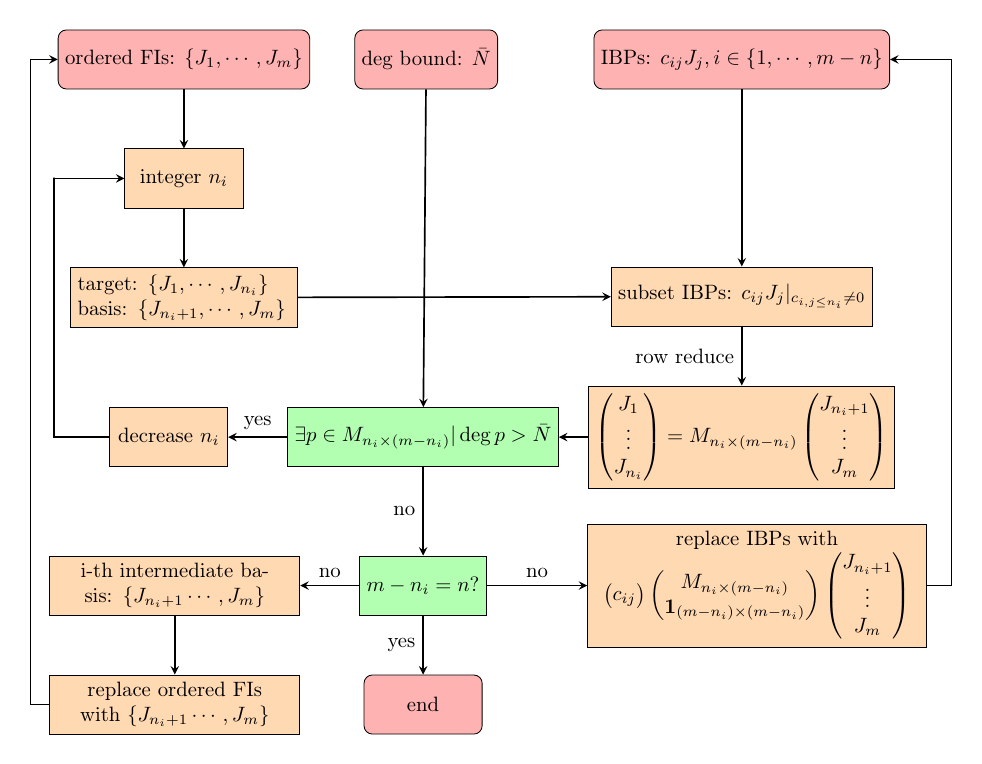}
    \caption{Flowchart of the algorithm for finding intermediate bases.}
    \label{fig:algorithm}
\end{figure}

The flowchart of the algorithm for finding intermediate bases is shown in Fig.\ref{fig:algorithm}. Once the intermediate bases are known, the reconstruction of basis transformation matrices is very straightforward, we therefore only focus on how to find intermediate bases. 

The algorithm starts with a complete set of IBP relations, the list of all Feynman integrals appearing in IBP relations, and a degree bound $\Bar{N}$. The IBP relations can be either symbolic in kinematic variables and $D$ or evaluated at numerical values. We use NeatIBP to generate all symbolic IBP relations, but one can also follow \cite{Page:2025gso,Smith:2025xes,delaCruz:2026mas,Anastasiou:2004vj,Lee:2012cn,Lee:2013mka,Smirnov:2008iw,Smirnov:2013dia,Smirnov:2014hma,Smirnov:2019qkx,Studerus:2009ye,vonManteuffel:2012np,Maierhofer:2017gsa,Klappert:2020nbg}. The Feynman integrals are ordered according to their reconstruction complexity as described in Sec.\ref{sec:intExample}, with master integrals put at the end. The degree bound $\Bar{N}$ serves as an upper bound on the rational functions in basis transformation matrices, and therefore controls the number of samples required for reconstruction. The exact value is a choice: a smaller $\Bar{N}$ typically leads to a larger value of $k$ of \cref{eq:mat}, meaning that more intermediate bases are needed to obtain simpler basis transformation matrices. The number of samples is fixed by $\Bar{N}=\mydeg{d_1,d_2}$ and the number of variables: a dense ansatz contains $\binom{|\vec{s}|+1+d_1}{d_1}+\binom{|\vec{s}|+1+d_2}{d_2}$ unknowns, independent of the problem. A larger $\Bar{N}$ reduces the number of
intermediate bases, but the corresponding linear systems grow rapidly, with the
cost of Gaussian elimination scaling roughly as the cube of their size.
$\Bar{N}$ is also bounded from below by the degrees of the coefficients in the
IBP relations. In practice, we choose the largest $\Bar{N}$ for which the
linear system can be solved with the available memory and reasonable
runtime; for the five- and six-variable problems considered here, this is
around $\mydeg{10,10}$, corresponding to $\mathcal{O}(10^4)$ unknowns.

The first step is to choose the first $n_i$ integrals as ``targets'' and the left integrals as ``basis''. After selecting relevant subset of IBP relations, reduction rules between ``targets'' and ``basis'' can be obtained via row reduction. We perform row reduction on multiple numerical points of kinematic variables and $D$, and derive the degree of all rational functions in the reduction rule. If the basis exchange matrix, $M_{n_i\times (m-n_i)}$, contains a rational function $r$ whose degree is higher than $\Bar{N}$, we decrease $n_i$ and repeat above steps. If no rational function has degree higher than $\Bar{N}$, we set ``basis'' as the $i$-th intermediate basis. The above steps takes very little time since row reduction is performed only on $n_i$ relations, typically $n_i=\mathcal{O}(10)$. The exact value of $n_i$ depends on the degree bound $\Bar{N}$: the smaller $\Bar{N}$ is, the smaller $n_i$ tends to be. 

The algorithm terminates when the number of integrals in the intermediate basis equals the number of the master integrals. At that point, all Feynman integrals have been reduced to master integrals according to \cref{eq:mat}. Otherwise, we proceed to the next iteration, replacing ordered Feynman integrals by the i-th intermediate basis and removing ``targets'' from IBP relations by applying reduction rules between ``targets'' and ``basis''.

Before closing this section, it is worth mentioning that the algorithm described above can also serve as an IBP reduction algorithm, similar idea was presented in \cite{Smith:2025xes}.

\section{Examples}\label{sec:example}
\begin{figure}
    \centering
    \includegraphics[width=\linewidth]{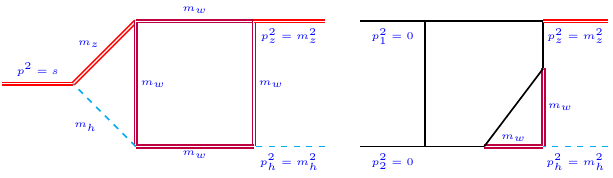}
    \caption{left: massive box-triangle; right: massive pentagon-triangle. The red, purple and blue line denote Z, W and Higgs, and black lines denote massless particles.}
    \label{fig:example2loop}
\end{figure}

\subsection{Massive Box-Triangle}
We use our method to reconstruct all reduction coefficients for a two-loop massive box-triangle, see Fig.\ref{fig:example2loop}. This topology contains the most complicated reduction coefficients among all vertex-like contributions to the two-loop electroweak corrections to $e^+e^-\to ZH$, and the degree of it is $\mydeg{35,36}$ in five variables $\{D,m_z^2,m_h^2,m_w^2,s\}$. The most general ansatz contains 1407406 terms, since this polynomial is not homogeneous due to $D$. The corresponding linear equation system required $\sim 15$Tb of memory, and solving it would takes $\sim 15$ years \footnote{Such an estimation is based on evaluation on 32-bit prime number with private linear solver.}. 

We set $\Bar{N}=\mydeg{9,8}$ and found 11 intermediate bases. Reconstructing all basis transformation matrices only needs 3289 samplings. The reduction result is cross-checked with NeatIBP at select numerical points over multiple different primes.

\subsection{Massive Pentagon-Triangle}
The second example considered is a two-loop massive pentagon-triangle, see Fig.\ref{fig:example2loop}. The most complicated reduction coefficients is of degree $\mydeg{40,42}$ in six variables $\{D,m_z^2,m_h^2,m_w^2,s,t\}$, and the most general ansatz contains 21638331 terms. We set $\Bar{N}=\mydeg{10,9}$, and find 13 intermediate bases. Reconstructing all entries of basis transformation matrices only needs 13013 samplings. The reduction result is cross-checked with NeatIBP on select numerical points over multiple different primes. 

We also test the numerical stability and efficiency of the matrices product at the following floating-point
\begin{gather}
m_z=91.1876,~m_h=125.1,~m_w=80.379,\\
s=240^2,t=1000
\end{gather}
while keeping $D$ symbolic. Evaluating the matrix products at the above numerical point takes $\sim 10$s. For comparison, we use Kira\cite{Maierhofer:2018gpa,Klappert:2020nbg,Maierhofer:2017gsa,Lange:2025fba} to reduce the same list of target integrals at the same numerical point, also keeping $D$ symbolic, which takes $\sim 66$s. Note that Kira returns reduction rules only for the target integrals, whereas the matrix product reduces all Feynman integrals appearing in the IBP relations generated by NeatIBP, which form a superset of the targets. Although this is not a direct benchmark, it indicates that, once the reconstruction is done, the reconstructed reduction coefficients are more efficient when one needs to evaluate them at a large number of kinematic points. 

To compare the two results, which are rational functions of $D$ with floating-point coefficients, we expand them around $D=4$ up to $\mathcal{O}(\epsilon^2)$ \footnote{For our interested part of physical amplitude, $\mathcal{O}(\epsilon^2)$ is enough}, where $\epsilon=(4-D)/2$. The relative difference of all reduction coefficients is below $10^{-10}$.

\section{Conclusions}\label{sec:conclusion}
In this paper, we present a new method for reconstructing the symbolic dependence of IBP reduction coefficients on the kinematic variables and the space-time dimension from  numerical IBP reduction. The basic idea is to iteratively express complicated Feynman integrals in terms of simpler Feynman integrals by row reducing a subset of IBP relations, until remaining Feynman integrals are the master integrals. The number of samplings needed for reconstruction is much smaller compared to the number of unknown in a general ansatz. The reduction coefficients are expressed as a product of several matrices, which enables fast and stable numerical evaluations.   

We have implemented this method to reconstruct the reduction coefficients for two non-trivial topologies which contributes to the two-loop electroweak corrections to $e^+e^-\to ZH$ process. This method is very efficient and flexible, and we expect it to be of crucial importance for perusing analytical result for scattering amplitude at higher loop orders.

\section*{Acknowledgments}
The author is grateful to Lisong Chen and Pavel P. Novichkov for providing feedback on this draft, and to Daniel Schieber for cross-checking the number of master integrals. The work of Qian Song was supported by the European Research Council (ERC) under the European Union’s Horizon Europe research and innovation program grant agreement 101078449 (ERC Starting Grant MultiScaleAmp). Views and opinions expressed are however those of the authors only and do not necessarily reflect those of the European Union or the European Research Council Executive Agency. Neither the European Union nor the granting authority can be held responsible for them.


%

\end{document}